\documentclass[
journal=nalefd,
manuscript=letter,
layout=twocolumn
]{achemso}

\usepackage{chemformula} 
\usepackage[T1]{fontenc} 
\usepackage{braket}
\usepackage{graphicx}
\usepackage{dcolumn}
\usepackage{bm}
\usepackage[normalem]{ulem} 
\usepackage{color}

\usepackage{hyperref}
\hypersetup{
    colorlinks,%
    citecolor=blue,%
    linkcolor=blue,%
    urlcolor=blue
}

\newcommand{\orcid}[1]{\href{https://orcid.org/#1}{\includegraphics[width=8pt]{orcid.pdf}}}

\author{F. Crasto de Lima}
\email{felipe.lima@lnnano.cnpem.br}
\author{A. Fazzio}
\email{adalberto.fazzio@lnnano.cnpem.br}
\affiliation{Brazilian Nanotechnology National Laboratory CNPEM, \\ C.P. 6192, 13083-970, Campinas, SP, Brazil}

\title{At the verge of topology: vacancy-driven quantum spin Hall in trivial insulators}


\begin{document}

\begin{tocentry}

\includegraphics[scale=1]{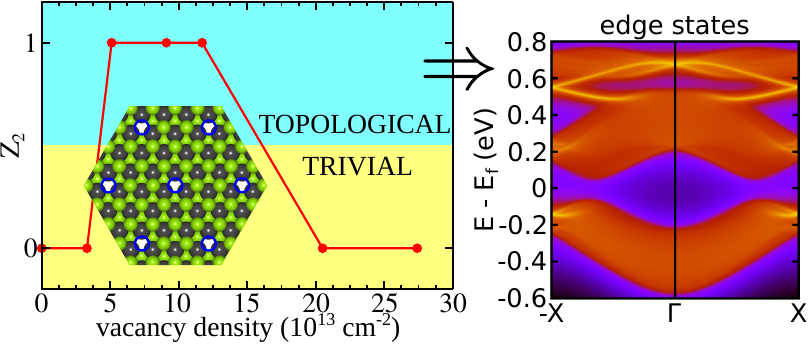}

\end{tocentry}

\begin{abstract}
Vacancies in materials structure -- lowering its atomic density -- take the system closer to the atomic limit, to which all systems are topologically trivial. Here we show a mechanism of mediated interaction between vacancies inducing a topologically non-trivial phase. Within an {\it ab initio} approach we explore topological transition dependence with the vacancy density in transition metal dichalcogenides. As a case of study, we focus on the PtSe$_2$, to which pristine form is a trivial semiconductor with an energy gap of $1.2$\,eV. The vacancies states lead to a large topological gap of $180$\,meV within the pristine system gap. We derive an effective model describing this topological phase in other transition metal dichalcogenide systems. The mechanism driving the topological phase allows the construction of backscattering protected metallic channels embedded in a semiconducting host.
\end{abstract}


The introduction of defects in materials has allowed extensive experimental modification of their properties. The most basic defect unit in materials is atoms vacancies. These localized defects in contrast with the more complex dislocations and grain boundaries can be directly induced and controlled experimentally \cite{PRcorbett1965, Rjiang2019}. Vacancies centers can act as active catalytic sites \cite{JACSli2016}, but also drive exotic effects, for instance, as charge density waves in In nanowires on Si(111) \cite{PRLpark2004} and ferromagnetism on transition metal dichalcogenides (TMD) \cite{JACScai2015}. Electronic phase transition (metal-insulator) have also taken place mediated by vacancy formations in GeSbTe \cite{SRbragaglia2016}.

A recent class of system presenting quantum topological transition have gained the interest in recent year. These systems present insulating behavior in their bulk but metallic surface states with symmetry stabilized backscattering protection. The defect-induced topological transition was observed in alloy systems by replacing atoms keeping within the same atomic density. For instance, in (Bi$_{1-x}$In$_{x}$)$_2$Se$_3$ a topological non-trivial to trivial insulating phase is achieved with an increase in the alloying parameter $x$ \cite{NLwang2019}, that is, increasing the defect concentration lead to the topological phase to vanishes. {Additionally, keeping the same atomic density, Anderson random disorder have also shown to drive a topological transition on HgTe quantum wells \cite{PRLli2009, PRLgroth2009}.} Besides substitutional defects, vacancy formation, i.e. reducing the atomic density, takes the systems towards the atomic limit to which all systems are trivial. 

In this paper, we show that another phenomena can take place, where the vacancy states give rise to a topological phase, that is, a trivial to non-trivial topological transition with the increase in defects concentration. Taking an {\it ab initio} approach based on the density functional theory (DFT) we investigate the vacancy driven topology in the trivial semiconductor PtSe$_2$. Extrapolating to general TMD systems, a simple model was derived to describe the threshold vacancy density for the topology to emerge. The vacancy states within the host bandgap allow topological states to be constructed within a semiconducting matrix.


Se vacancy on PtSe$_2$ is the most energetically favorable defect \cite{PRBabsor2017}. Experimental Se vacancy formation on PtSe$_2$ can be controlled by electron irradiation \cite{PRLkomsa2012}, while its assembly can be achieved by a thermal treatment \cite{NATMATlin2017} by the mobility of such defects \cite{ASCOMEGAgao2017}. Additionally, with the recent advances on controlled atomic positioning on surfaces by AFM/STM tips \cite{NATNANOpavlicek2017, NATPHYSslot2017, NLcrasto2020}, controlled Se vacancies have been induced on the TMD PdSe$_2$ \cite{PRLnguyen2018}.

\begin{figure}[h!]
\includegraphics[width=\columnwidth]{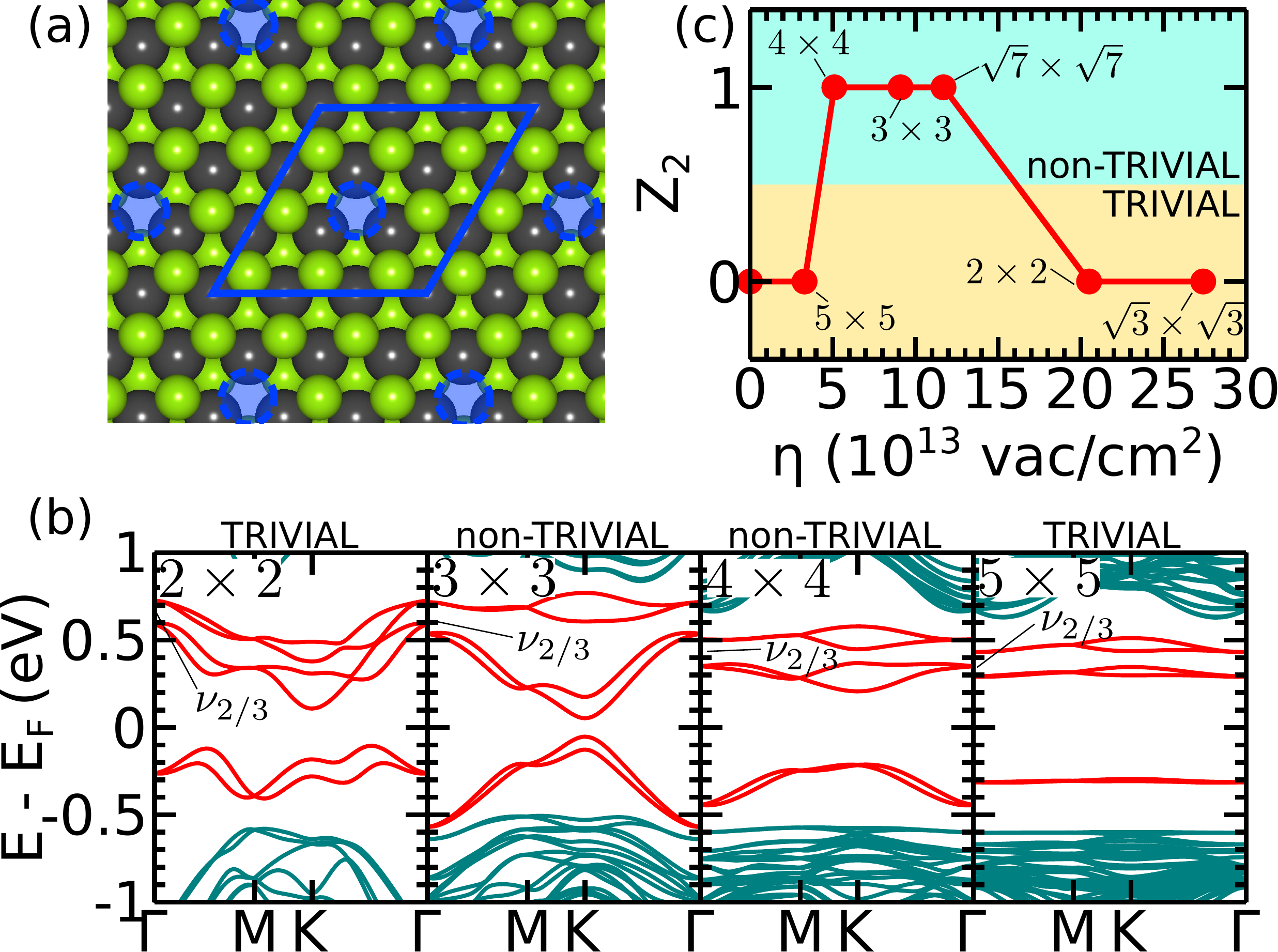}
\caption{\label{hex-geo} (a) $3\times3$ supercell with the Se vacancies geometry. (b) DFT band structure {with SOC} for hexagonal cells with vacancy concentrations from $8.2\times 10^{14}$\,cm$^{-2}$ down to $5.1\times 10^{13}$\,cm$^{-2}${, the blue (red) bands are the host (vacancy) states.} (c) $Z_2$ invariant {at $\nu_{2/3}$ filing} for different Se vacancy concentrations.}
\end{figure}

From the electronic perspective, an isolated Se vacancy introduces localized states in the host PtSe$_2$ bandgap. Upon the removal of a Se atom, three lone pairs arise in the Pt atoms neighboring the vacancy. By increasing the vacancy density, those states can interact {with each other} forming energy bands. Indeed the in-gap vacancy states, have a major contribution from Pt $d_{xz}$/$d_{yz}$, with its interaction mediated by the host Se $p$ orbitals \cite{PRBabsor2017}. Such a picture of interacting vacancy states within the host energy gap, ruling a hopping mediated charge transport, is experimentally observed in TMD \cite{NATCOMqiu2013}. The dispersion of those in-gap vacancy states (ruled by the vacancy concentration), allied with the host TMD spin-orbit coupling (SOC) has the ingredients for a possible topologically non-trivial phase. {In Figure \ref{hex-geo}(a) we show a single vacancy on a PtSe$_2$ 3$\times$3 supercell, taking different supercell sizes defines the Se vacancy concentration on this system.} { In Figure \ref{hex-geo}(b) we shown the DFT calculated band structure, with the red bands being the vacancy states neatly laying within the host gap states (blue bands).} We see that the in-gap states (red bands) are dispersive for higher vacancy densities [Figure \ref{hex-geo}(b)] (smaller supercell size), with its dispersion getting narrower with decreasing vacancy density. It is worth pointing out that the discussed Se-vacancy densities are within the values achieved experimentally in PtSe$_2$, up to $2.5\times10^{14}$ vac/cm$^{2}$ (PtSe$_{1.75}$) \cite{NATMATlin2017}, with observed chalcogenides monovacancies in other TMD reaching up to $5.2\times10^{14}$ vac/cm$^2$ (MoS$_{1.54}$) \cite{NATMATli2016}.

{The spin-orbit coupling effect within the vacancy states can lead to topological phases.} Indeed we have calculated the $Z_2$ invariant for the two energy gaps of the vacancy states, at $1/3$ filing (at the Fermi energy), and $2/3$ filing (above the Fermi energy). At $1/3$ filing all systems are trivial, however at $2/3$ we found a topological transition with the vacancy density increase [Figure \ref{hex-geo}(c)]. The system with zero vacancies ($\eta=0$), pristine PtSe$_2$, is a trivial semiconductor with a $1.2$\,eV bandgap. Introducing low vacancy densities, $\eta < 3\times10^{13}$vac/cm$^2$, the system is still trivial as the vacancies states are localized and do not interact with each other. Increasing the vacancy density such scenario change and the interaction between the vacancy states leads to a topological phase up to $\eta = 13\times10^{13}$vac/cm$^2$ (PtSe$_{1.94}$). The electron doping required to access the $2/3$ filing topological gap is $2\eta$ which is accessible in 2D systems by gating interfaces \cite{AMliang2020}. Further increasing of the vacancy density leads to a re-entrant trivial phase. {It is worth pointing out that we have introduced the vacancies in other crystalline cell geometries, for instance in a tetragonal cell, and the same behavior is observed concerning the topological phase and vacancy concentration. Additionally, 2D topology was recently characterized as robust against disordered assemblies \cite{2DMfocassio2021, PRMpezo2021}.}


\begin{figure}[h!]
\includegraphics[width=\columnwidth]{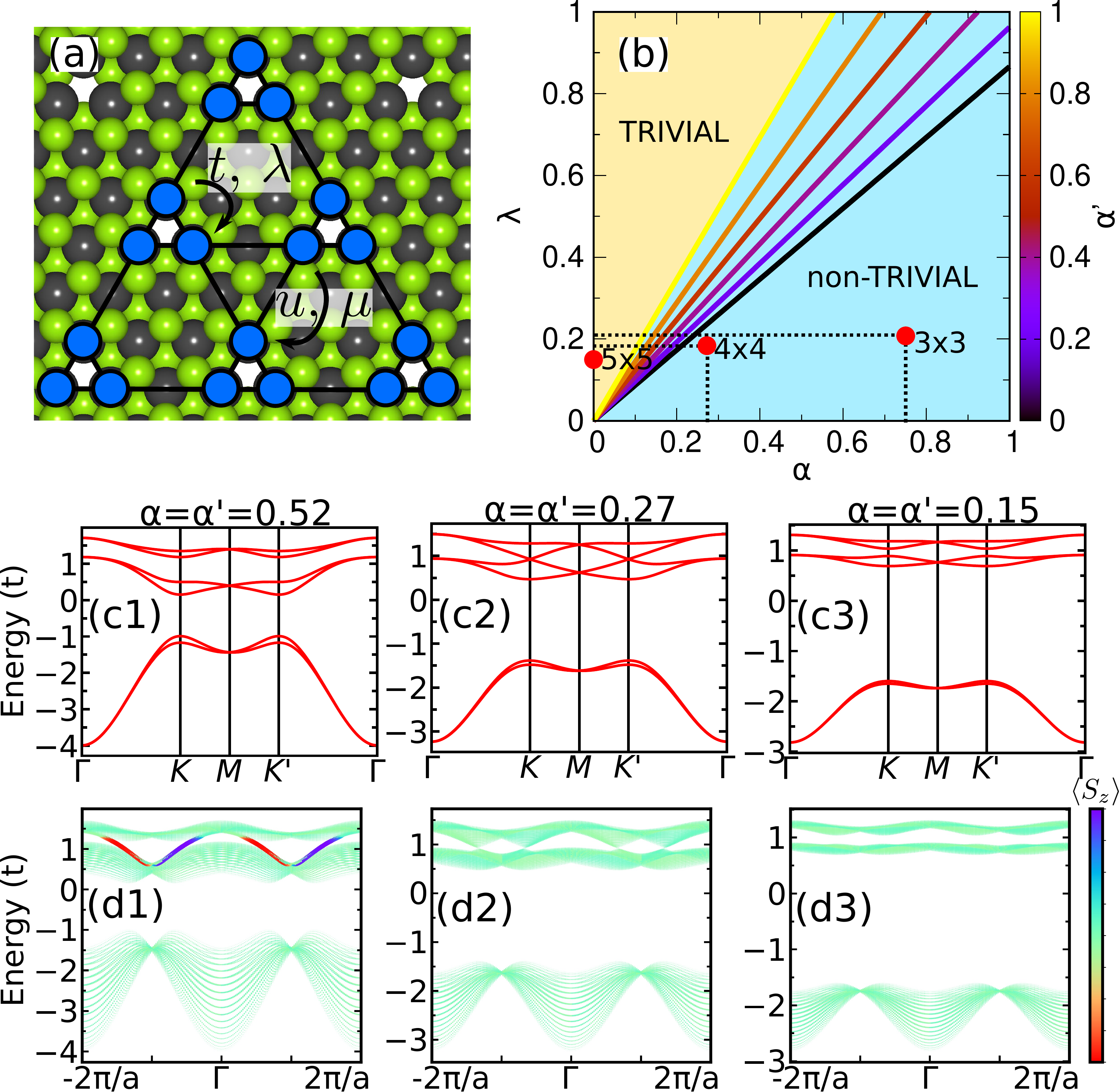}
\caption{\label{phase-diag} (a) Tight-binding model geometry. (b) Topological phase diagram at $2/3$ filing for the hexagonal cell{, the colored diagonal lines dictates the phase boundary for different SOC attenuation factor ($\alpha'$), while the blue/yellow region mark the non-trivial/trivial parameter space}. (c1)-(c3) Bulk bands and (d1)-(d3) edge states evolution as a function of $\alpha$ for $t=-1.31$ and $\lambda=-0.1\,t$.}
\end{figure}

\begin{figure*}
\includegraphics[width=1.8\columnwidth]{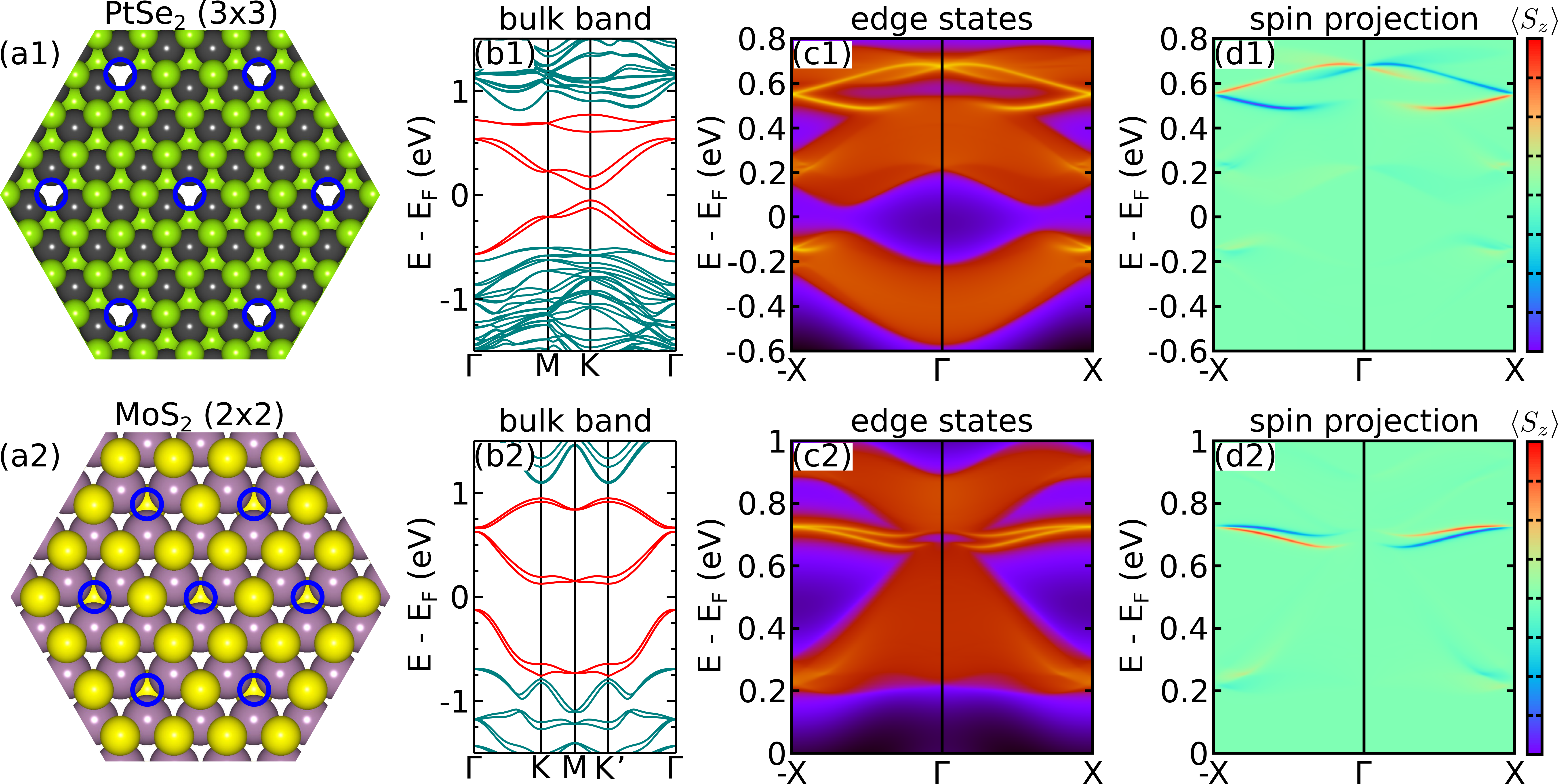}
\caption{\label{mos2} (a) Vacancy structure, (b) bulk band structure, (c) semi-infinite edge states, and (d) spin projected edge states. PtSe$_2$ ($3\times3$) supercell in (a1)-(d1) and MoS$_2$ ($2\times2$) supercell in (a2)-(d2).}
\end{figure*}

To understand the interplay between the vacancy concentration and the emergence of the non-trivial state, we have extracted a model for the phase transition in the hexagonal geometry. The hamiltonian for the three lone pairs interacting in the presence of SOC is written as
\begin{equation}
H = \sum_{\langle ij \rangle} (t\sigma_0 + i\lambda v_{ij} \sigma_z) c_i^\dagger c_j  + \sum_{\langle \langle ij \rangle \rangle} (u\sigma_0 + i \mu v_{ij} \sigma_z) c_i^\dagger c_j
\end{equation}
where $t$ and $u$ describe the hopping between the nearest and next-nearest neighbors sites while $\lambda$ and $\mu$ the SOC induced term [as shown in Figure \ref{phase-diag}(a)]; $\sigma_i$ are the spin Pauli matrix{, with $\sigma_0$ the 2$\times$2 identity matrix,} and $v_{ij}$ the sign of the SOC contribution \cite{PRBdelima2017}. Such Hamiltonian assumes the form
\begin{equation}
H(\vec{k})=
\begin{pmatrix}
0 & 0 & A_{\vec{k}} & 0 & B_{\vec{k}} & 0 \\
0 & 0 & 0 & A^*_{-\vec{k}} & 0 & B^*_{-\vec{k}} \\
A^*_{\vec{k}} & 0 & 0 & 0 & C_{\vec{k}} & 0 \\
 0 & A_{-\vec{k}} & 0 & 0 & 0 & C^*_{-\vec{k}} \\
B^*_{\vec{k}} & 0 & C^*_{\vec{k}} & 0 & 0 & 0 \\
0 & B_{-\vec{k}} & 0 & C_{-\vec{k}} & 0 & 0 
\end{pmatrix}
\end{equation}
with
\begin{eqnarray}
A_{\vec{k}} &=& t+i\lambda + (u+i\mu) e^{i\vec{k}\cdot \vec{a}_2}; \\
B_{\vec{k}} &=& t-i\lambda + (u-i\mu) e^{i\vec{k} \cdot (\vec{a}_2 - \vec{a}_1)}; \\
C_{\vec{k}} &=& t+i\lambda + (u+i\mu) e^{-i\vec{k} \cdot \vec{a}_1}.
\end{eqnarray}
Given the preservation of time-reversal, at the $\Gamma$ point all eigenvalues are double degenerated with energies
\begin{eqnarray}
E_{1,0}^{\Gamma} &=& 2t+2u, \\
E_{2,\pm}^{\Gamma} &=& \pm \sqrt{3} (\lambda + \mu) - (t+u),
\end{eqnarray}
while at the K point the SOC terms break the double degeneracy with
\begin{eqnarray}
E_{1,\pm}^K &=& \pm \sqrt{3} \lambda - t + 2u, \\
E_{2,\pm}^K &=& \pm \sqrt{3}(\lambda - \mu) - (t+u), \\
E_{3,\pm}^K &=& \pm \sqrt{3} \mu + 2t -u .
\end{eqnarray}

We can write $u=\alpha t$ and $\mu=\alpha' \lambda$ with the attenuation coefficients ($\alpha$, $\alpha'$) capturing the different hopping distances between lone pairs of the same vacant site, and neighboring vacant site. This attenuation coefficient is directly connected with the vacancy density, i.e. for higher vacancy density $\alpha \rightarrow 1$ while for lower density $\alpha \rightarrow 0$. Those parameters can be directly extracted from the DFT calculations for instance comparing the band-width at $\Gamma$ ($W_{\Gamma}$), the gap at K in the $1/3$ filling ($G_K$), the gap at $\Gamma$ in the $2/3$ filling ($G_{\Gamma}$) and the SOC split at K of the topmost bands ($S_{K}$) being
\begin{eqnarray}
\alpha &=& \frac{W_{\Gamma} - G_{K}}{W_{\Gamma} + G_{K}}; \;\; t = \frac{W_{\Gamma}}{3(1+\alpha)}; \\
\alpha' &=& \frac{G_{\Gamma} - S_{K}}{G_{\Gamma} + S_{K}};\;\; \lambda = \frac{G_{\Gamma}}{3(1+\alpha')}.
\end{eqnarray}

The topological phase boundary at $2/3$ filing can be extracted from the gap closing at the K point given by {the relation $\lambda = \sqrt{3} \alpha/(2-\alpha')$, such curve in the parameter space is depicted by the diagonal colored lines in Figure \ref{phase-diag} (b)}. Here, for a given value of NN hoping and SOC ($t$ and $\lambda$), which is material dependent, the $\alpha$ decaying factor (associated with the vacancy concentration) define the trivial/topological phase. For instance, taking $t=-1.31$ and $\lambda = -0.1$\,$t$ we show the {tight-binding} band structure and spin-projected edge states for $\alpha=\alpha'$ going from $0.52$ to $0.15$ [Figure \ref{phase-diag}(c1)-(d3)]. Here, {the model band structure are the vacancy red bands appearing on the DFT calculations of Figure \ref{hex-geo}(b). F}or $\alpha=0.52$ the system resembles the characteristic kagome bands of a triangular assembly [Figure  \ref{phase-diag}(c1)], with the edge states appearing at $2/3$ filing [Figure \ref{phase-diag}(d1)]. By decreasing the decay factor to $\alpha=0.27$, that is, lowering the vacancy concentration, the gap at $2/3$ filing is closed and the edge states vanish [Figure \ref{phase-diag}(c2) and (d2)]. Further decreasing the vacancy concentration, $\alpha = 0.15$, the gap reopens, but with a trivial character, and the edge states are no long present [Figure \ref{phase-diag}(c3) and (d3)]. As a validation of the proposed model we have located the parameters for PtSe$_2$ DFT simulation at the phase diagram [Figure \ref{phase-diag}(b)]. This tight-binding model is useful to capture the topological phase of the system agreeing with the {\it ab initio} topological invariant ($Z_2$) calculation, and allow for an interpretation of required vacancy concentration for topology emergence in other TMD systems.

The topological phase in the PtSe$_2$ vacancy system is quite robust, allowing experimental observation. Particularly, for the $3 \times 3$ vacant hexagonal cell $9.1\times10^{13}$\,vac/cm$^2$ (PtSe$_{1.89}$) the topological SOC energy gap at $\Gamma$ is $180$\,meV. Such values are higher than most predicted quantum spin Hall systems \cite{NLmarrazzo2019}, and as we will show below, this guaranteed for a ribbon of only $4.3$\,nm to already present the characteristic topological edge states. To highlight the topological phase we have constructed a semi-infinite system and characterize the edge states, as shown in Figure \ref{mos2}. Here, we can observe the edge states appearing within the bulk energy at $2/3$ filing, Figure \ref{mos2}(c1), with the spin-texture of such edge states obeying the backscattering forbidden character, Figure \ref{mos2}(d1). S-vacancy formation is also achieved in MoS$_2$ system \cite{NATCOMhong2015, NLlu2014} up to $5.2\times10^{14}$\,vac/cm$^2$ \cite{NATMATli2016}. Here a similar vacancy-driven topological mechanism is observed [Figure \ref{mos2}(a2)-(d2)], however with a smaller topological gap of $40$\,meV for $\eta =2.8 \times 10^{14}$\,vac/cm$^2$, given the lower relativistic effect of the Mo atoms.

\begin{figure}[h!]
\includegraphics[width=\columnwidth]{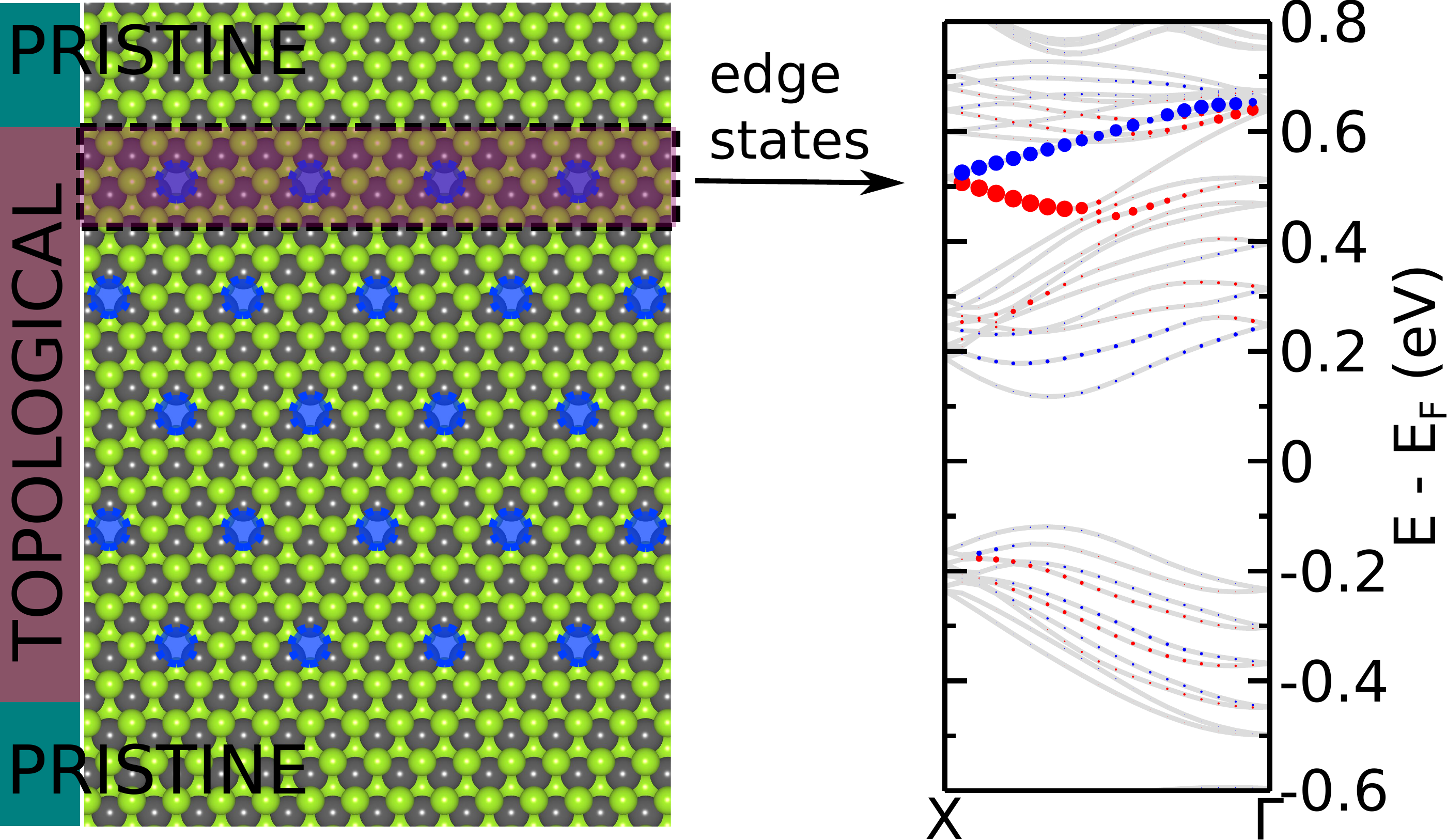}
\caption{\label{interface} Topological system driven by vacancy embedded on PtSe$_2$ semiconducting host. The edge states are well distinguished from the bulk bands. Blue/red circles size (right panel) are proportional to the edge states spin up/down contribution to the band structure.}
\end{figure}

Taking advantage of the vacancy-controlled assembly by thermal treatment \cite{NATMATlin2017} in TMD, we have shown that the generation of topological states embedded on the semiconducting 2D systems can be constructed. Here we explicit construct the interface between pristine PtSe$_2$ and a Se vacant region within DFT calculations [Figure \ref{interface}]. We took a nanoribbon with a non-trivial region (vacant region) of $4.3$\,nm separated from its periodic images by a trivial region, pristine PtSe$_2$. The projected edge states' contribution to the band structure (circles in the Figure \ref{interface} band) further confirms the characteristic spin-polarized states arising at the interface.

In conclusion, we report on the topological phase driven by vacancies on transition metal dichalcogenides. We show that the mediated interaction between the vacancies, together with the host SOC, leads to non-trivial states neatly lying on the semiconductor host energy gap. As a case of study, we characterize by {\it ab initio} approach the topological states in PtSe$_2$, showing also the emergence on MoS$_2$. In the former a topological gap of $180$\,meV is among the highest predicted so far. A model enlightening the vacancy density dependence of the topological transition was obtained, which was shown to be useful in the prediction of the vacancy threshold for the topological phase. The proposed mechanism of vacancy-driven topology, allied with recent developments on autonomous control of atoms/vacancies on surfaces could allow an imprint of topological circuits in 2D systems in future device design.

\section*{Computational Approach}
Density functional theory calculations were performed within the Vienna {\it Ab initio} Simulation package \cite{CMSkresse1996}, with the PBE exchange-correlation term \cite{PRLperdew1996}, a $400$\,eV plane-wave base cutoff. The electron-ion interaction is considered within the projected augmented wave method \cite{PRBblochl1994}, with all atoms relaxed until forces were lower than $10^{-2}$\,eV/{\AA}. The special k-point of the 2D Brillouin zone was taken as $7 \times 7$ Monkhorst-Pack grid for the $1\times1$ cell and the same k-point density was used for the other cells. The relativistic SOC effect was considered in all calculations. The $Z_2$ invariant was calculated by the Wannier base extraction for the vacancy states within the wannier90 package \cite{JPCMgiovanni2020}.

\begin{acknowledgement}
The authors acknowledge financial support from the Brazilian agencies FAPESP (grants 19/20857-0 and 17/02317-2), INCT-Nanomateriais de Carbono,, and Laborat\'{o}rio Nacional de Computa\c{c}\~{a}o Cient\'{i}fica for computer time.

\end{acknowledgement}

%
%

\bibliography{bib}

\end{document}